\documentclass[aps,preprint,pra,groupedaddress,amsmath,amssymb]{revtex4-2}

\usepackage{hyperref}
\usepackage{graphicx}
\usepackage{dcolumn}
\usepackage{bm}

\begin{document}


\title{Near-wall lubricating layer in drag-reduced flows of rigid polymers}

\author{Lucas Warwaruk}
\affiliation{Department of Mechanical Engineering, University of Alberta, Edmonton, Alberta T6G 1H9, Canada}

\author{Sina Ghaemi}
\email[]{ghaemi@ualberta.ca}
\affiliation{Department of Mechanical Engineering, University of Alberta, Edmonton, Alberta T6G 1H9, Canada}

\date{\today}

\begin{abstract}
The current theories on the mechanism for polymer drag-reduction (DR) are generally applicable for long-chain flexible polymers that form viscoelastic solutions. Rigid polymer solutions that generate DR seemingly lack prevalent viscoelastic characteristics. They do, however, demonstrate larger viscosities and a noticeable shear-thinning trend, well approximated by generalized Newtonian models. The following experimental investigation scrutinized the flow statistics of an aqueous xanthan gum solution in a turbulent channel flow, with friction Reynolds numbers ($Re_{\tau}$) between 160 and 680. The amount of DR varied insignificantly between 28\% and 33\%. The velocity field was measured using planar particle image velocimetry and the steady shear rheology was measured using a torsional rheometer. The results were used to characterize the flow statistics of the polymer drag-reduced flows at different $Re_{\tau}$ and with negligible changes in DR; a parametric study only previously considered by numerical simulations. Changes to the mean velocity and Reynolds stress profiles with increasing $Re_{\tau}$ were similar to the modifications observed in Newtonian turbulence. Specifically, the inner-normalized mean velocity profiles overlapped for different $Re_{\tau}$ and the Reynolds stresses monotonically grew in magnitude with increasing $Re_{\tau}$. Profiles of mean viscosity with respect to the wall-normal position demonstrated a thin layer that consists of a low-viscosity fluid in the immediate vicinity of the wall. Fluid outside of this thin layer had a significantly larger viscosity. We surmise that the demarcation in the shear viscosity between the inner ``lubricating" layer and the outer layer cultivates fluid slippage in the buffer layer and an upward shift in the logarithmic layer; a hypothesis akin to DR using wall lubrication and superhydrophobic surfaces.
\end{abstract}

\maketitle


\newpage

\section{Introduction}

Adding small quantities of high molecular weight polymers to a turbulent wall flow can produce a large reduction in skin friction relative to the solvent alone. This phenomenon was first discovered by Toms in 1948 \cite{Toms1948} and has since been actively studied. The most readily used drag-reducing polymers are long-chain flexible molecules, that form viscoelastic non-Newtonian solutions. When these flexible polymer solutions are subjected to large amounts of shear (for example through a pump, fitting or a restriction) the solution undergoes mechanical degradation, and becomes less effective at mitigating drag \cite{DenToonder1995}. Therefore, flexible polymers are often used as a method for cost-saving or performance enhancement in once-through fluid transport systems such as crude oil pipelines, fire-suppression, or municipal sewage \cite{Burger1980,Sellin1980,Figueredo2003}. Mechanical degradation can be reduced by utilizing a polymer with a rigid molecular structure; however, this may require higher additive concentrations and a sacrifice in the amount of drag-reduction (DR) \cite{Pereira2013}. While flexible polymers may coil and stretch within the turbulent flow, rigid polymers are believed to remain elongated at all times, regardless of the imposing flow field \cite{Bewersdorff1988}. Rigid drag-reducing polymers are often naturally-occurring polysaccharides that come in many different variants \cite{Bewersdorff1988,Japper-Jaafar2009,Pereira2013,Soares2019,DosSantos2020}. While DR using flexible polymers has been readily studied, investigations of rigid polymer DR are much less abundant. Existing comparisons seem to imply flexible and rigid polymers mitigate drag in entirely unique mannerisms \cite{Virk1990,Mohammadtabar2017,Warwaruk2021}. However, additional investigations are needed to support such a conclusion. The following overview details the previous theoretical and experimental findings pertinent to polymer DR using flexible and rigid molecules.

Two existing theories attempt to elucidate how polymers interact with turbulence and reduce drag. In Lumley's ``viscous theory" of DR, polymers are believed to damp turbulent structures due to an enhanced extensional viscosity in regions of the flow that experience significant elongational deformation rates \cite{Lumley1973}. The cogency of Lumley's viscous theory is conditional on polymer solutions having a large Trouton ratio, $Tr = \mu_{ext}/\mu > 3$, and Weissenberg number, $Wi = \lambda_{ext} / \tau_{f} > 1/2$. Here $\mu_{ext}$ is the extensional viscosity, $\mu$ is the shear viscosity, $\lambda_{ext}$ is the extensional relaxation time, and $\tau_{f}$ is a representative time scale of the turbulent flow. The ``elastic theory," proposed by de Gennes \cite{DeGennes1990}, suggested that a large $\mu_{ext}$ is not critical for DR. Instead DR occurs when the elastic stresses of the polymers become comparable to the Reynolds stresses of the flow. Regardless of the validity of one theory over the other, both rely on the polymer solution having some amount of viscoelasticity. This is despite the insinuation that the name of Lumley's ``viscous theory" may prescribe -- a caveat that has been alluded to by other authors as well \cite{White2008,Xi2019}. Indeed, the results of pipe and channel flow experiments using flexible polymers have demonstrated positive correlation between the amount of DR and the \textit{Wi} of the flow \cite{Owolabi2017}. Also, investigations using direct numerical simulation (DNS) with viscoelastic constitutive equations, such as the molecular FENE-P model (finitely extensible nonlinear elastic dumbbell model with a Peterlin approximation) and the continuum-based Oldroyd-B model, have demonstrated good agreement in trends of skin-friction, mean velocity and Reynolds stresses, when compared with experimental results using flexible polymers \cite{Ptasinski2003,Min2003a,Min2003b}. All analytical, experimental and numerical evidence leads us to believe that viscoelasticity is a necessity for polymer DR. However, rheological measurements of dilute rigid polymer solutions exhibit little extensional and elastic characteristics, despite being able to produce DR similar to flexible polymer solutions \cite{Escudier2009,Pereira2013,Warwaruk2021}. Comparisons of each solutions rheology can directly quantify the unique material characteristics of the two solutions.

Numerous experimental investigations have measured the shear and extensional rheology of flexible and rigid polymer solutions \cite{Owolabi2017,Mohammadtabar2020,Warwaruk2021}. Mohammadtabar et al. \cite{Mohammadtabar2020} directly compared $\mu_{ext}$ of solutions of different drag-reducing flexible and rigid polymer species using a capillary breakup extensional rheometer (CaBER). Their results reflected that drag-reducing flexible polymer solutions, with concentrations as low as 20 ppm, can exhibit $Tr$ as large as 100 \cite{Mohammadtabar2020}. The implication is that flexible polymer solutions are viscoelastic ($Tr > 3$) and effective at resisting elongational strain-rates \cite{Barnes1989}. For the rigid polymer solutions, Mohammadtabar et al. \cite{Mohammadtabar2020}, found that the extensional properties of the solution were immeasurable using the CaBER apparatus; an observation not uncommon among experimentalists \cite{Escudier2009,Japper-Jaafar2009,Mohammadtabar2020,Warwaruk2021}. Attempts at measuring $\mu_{ext}$ for rigid polymer solutions using a CaBER often fail, potentially due to a low $\mu_{ext}$ and $Tr$ \cite{Escudier2009,Mohammadtabar2020,Warwaruk2021}. However, the lack of extensional resistance in dilute rigid polymer solutions appears to be substituted with a larger shear viscosity and a more pronounced shear-thinning quality, when contrasted with solutions of flexible polymers that produce similar quantities of DR \cite{Escudier2009,Mohammadtabar2020,Warwaruk2021}. Shear viscosity measurements of rigid polymer solutions are well approximated by shear-thinning generalized Newtonian (GN) models such as the Carreau-Yasuda or Cross models \cite{Carreau1972, Yasuda1981, Barnes1989}. Based on existing measurements, a well-defined shear-thinning trend is the only obvious rheological trait for low-concentration solutions of drag-reducing rigid polymers \cite{Mohammadtabar2020,Warwaruk2021}. The rheological measurements of flexible and rigid polymer solutions suggest that the two additives have different mechanisms for promoting DR.

Further evidence for a unique DR mechanism among flexible and rigid polymers is demonstrated by differences in some of their flow statistics. The most apparent distinction is their unique trajectories for attaining the maximum drag-reduction (MDR) asymptote in skin-friction coefficient, $C_f$, with increasing polymer concentration or Reynolds number (Re). Virk \& Wagger \cite{Virk1990} defined flexible additives as Type A drag-reducers and rigid additives as being Type B, based on their different trends in $C_f$ versus Re. Despite their unique trends in $C_f$, flexible and rigid polymers share similarities in their mean velocity profiles. Escudier et al. \cite{Escudier2009} and Mohammadtabar et al. \cite{Mohammadtabar2017} demonstrated that mean velocity profiles of drag-reduced channel flows using rigid polymers were consistent with the elastic sublayer model, derived using mostly flexible polymers by Virk \cite{Virk1971}. On the other hand, some authors have suggested that the two solutions modify the Reynolds stresses differently. In a recent review publication, Xi \cite{Xi2019} stated that rigid polymers and flexible polymers produce different changes to the streamwise Reynolds stress profile during the transition from low drag-reduction (LDR) to high drag-reduction (HDR). Xi \cite{Xi2019} made this observation based on the same measurements by Escudier et al. \cite{Escudier2009} and Mohammadtabar et al. \cite{Mohammadtabar2017}. However, Xi \cite{Xi2019} also noted that a good portion of the modifications to the Reynolds stresses could be attributed to differences in Re, something demonstrated explicitly by Thais et al. \cite{Thais2012} using DNS and FENE-P, but seldom explored experimentally. Warwaruk \& Ghaemi \cite{Warwaruk2021} compared the Reynolds stresses of flexible and rigid polymers for flows at HDR and MDR. They observed inconsistencies in the Reynolds stresses for drag-reduced flows of similar DR, and concluded that the differences are primarily attributed to discrepancies in Re. Generally, it is not sufficiently well understood how the first- and second-order flow statistics of rigid polymer flows depend on Re and DR \cite{Escudier2009, White2008}. Few, if any, investigations have explicitly measured changes in the mean velocity profile and Reynolds stresses of drag-reducing rigid polymer solutions with varying Re that are independent of variations in DR. 

Although the two most cited theories for polymer DR rely on viscoelasticity being a constituent property of the polymer solution, there is one phenomenological model that simulates polymer DR using a viscous approximation. L'Vov et al. \cite{Lvov2004}, De Angelis et al. \cite{DeAngelis2004} and Procaccia et al. \cite{Procaccia2008}, demonstrated that polymer DR can be approximated by a simulation of the Navier-Stokes equations using an effective viscosity that is a function of the wall-normal distance from the wall, i.e. \textit{y}. Their effective viscosity profile was a piece-wise function, where in the linear viscous sublayer the effective viscosity was constant, but in the buffer and log layers the effective viscosity grew linearly with increasing \textit{y}. L'Vov et al. \cite{Lvov2004} and De Angelis et al. \cite{DeAngelis2004} derived and simulated the model based on simplifications done to the viscoelastic FENE-P constitutive equations; a model noted as being representative of flexible polymers. Procaccia et al. \cite{Procaccia2008} expressed that such an effective viscosity model could also be viable for modeling the drag-reduced flow of rigid polymer solutions as well. Instead of using FENE-P, Procaccia et al. \cite{Procaccia2008} derived the effective viscosity model for rigid polymers based on a constitutive equation derived by Doi \& Edwards \cite{Doi1988} for dilute solutions of rod-like molecules. However, given the innate shear-thinning quality of rigid polymer solutions, coupled with the mean wall-normal velocity gradient in a turbulent wall flow, an effective viscosity that increases with respect to the wall-normal distance is inherent in the turbulent wall flow of rigid polymer solutions. Indeed, DNS using GN shear-thinning constitutive equations demonstrates a spatially varying mean viscosity profile, where the viscosity was relatively constant near the solid boundary, but increased logarithmically in the buffer and log layers along \textit{y} \cite{Singh2017,Singh2018,Arosemena2020,Arosemena2021}. Although the wall-normal trend in the mean viscosity for GN fluids is different than the linear effective viscosity profile used in De Angelis et al. \cite{DeAngelis2004}, it is clear that wall-normal variations in the viscosity may play an important role in DR. Therefore, an experimental investigation that evaluates the wall-normal viscosity profile of a rigid polymer solution is warranted.

The present investigation has two central objectives. First, to provide high fidelity turbulence statistics of a drag-reduced channel flow of rigid polymers with varying Re. Few experiments of rigid polymers have explored the effect of Re on flow statistics. The existing measurements of rigid polymers in a turbulent channel flow have low spatial resolutions \cite{Escudier2009} or appear to be in an arguably transitional flow regime due to small Re \cite{Mohammadtabar2017}. To alleviate this gap in the research, an experimental investigation is performed for a 170 ppm xanthan gum (XG) solution in a turbulent channel flow with a friction Reynolds number, $Re_{\tau}$, between 160 and 680. The resulting levels of DR are between 28-33\% (i.e. LDR flows), demonstrating little dependence on Re. Planar particle image velocimetry (PIV) measurements are used to measure the instantaneous velocity of the drag-reduced flows. Shear rheology is characterized using a double gap and a parallel plate geometry to capture the viscosity of the XG solution over a large range of shear rates. The second objective of our manuscript is to elucidate a mechanism for rigid polymer DR from the perspective of lubricated flows. We observe a thin layer of low viscosity fluid near the wall for the rigid polymer solution at all flow conditions, which is proposed to be essential for DR using rigid polymers.

\section{Experimental methodology}

\subsection{Flow facility}\label{sec:2.1}

\begin{figure}[b]
	\includegraphics{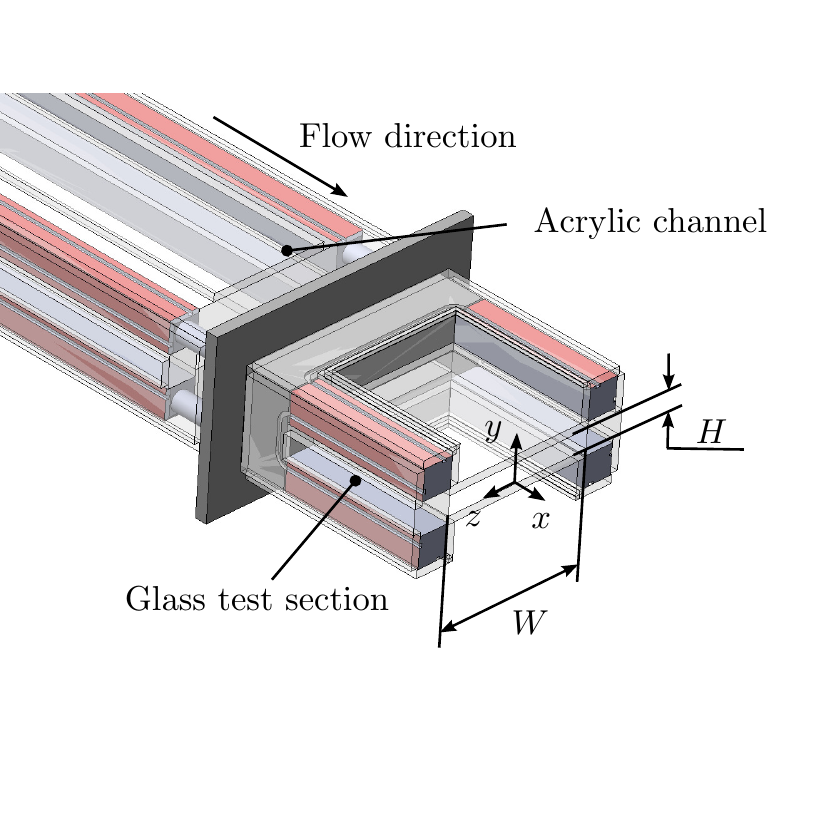}
	\caption{\label{fig:1} Isometric cross-section of the glass test section.}
\end{figure}

Experiments were performed in a recirculating flow facility with a channel section dedicated for flow measurements. A portion of the loop had a rectangular cross-section of height, $H$, of 15 mm and width, $W$, of 120 mm, as shown in figure \ref{fig:1}. Walls of the channel were cast acrylic, with the exception of one segment, the test section, where the walls were glass. Flow measurements using PIV were performed in the glass test section, which was situated approximately $107H$ downstream of the inlet to the channel portion. In total, the channel portion of the loop was $168H$ in length. Gradual transition fittings were used to connect the channel section to the remainder of the loop, which was 2 inch nominal pipe. Additional details pertaining to the flow facility can be found in Warwaruk \& Ghaemi \cite{Warwaruk2021}. Figure \ref{fig:1} displays the Cartesian coordinate system with reference to the cross-section of the glass test section. The standard right-hand orthonormal basis was used, where position along the streamwise, wall-normal and spanwise directions were denoted as $x$, $y$, and $z$, respectively. The coordinate system was placed at the mid-span of the lower channel wall within the laser sheet.

A centrifugal pump (LCC-M 50-230, GIW Industries Inc.) and a variable frequency drive (VFD) were used to propel the fluid within the loop. A shell and tube heat exchanger and a thermocouple were used to maintain a constant fluid temperature of 25$\pm 0.3 ^{o}$C. A Coriolis flow meter (Micro Motion F-series, Emerson Process Management) with an accuracy of $\pm$0.2\% was used to measure the mass flow rate, $\dot{m}$, of the moving fluid. To maintain a constant $\dot{m}$, a proportional-integral-derivative controller, developed using LabView software (LabView 2015, National Instruments), was used to manipulate the frequency of the VFD, and the rotational speed of the pump. The linear streamwise gradient in static pressure, $\Delta P /\Delta x$, was measured using a differential pressure transducer (DP15, Validyne) equipped with a 1 psi diaphragm. Pressure ports were separated by $ \Delta x = 109H$. The upstream port was situated $34H$ from the inlet of the channel.

\begin{table*}[t]
	\begin{ruledtabular}
		\centering
		\caption{\label{tab:1}Flow properties for channel flow of water.}
		\begin{tabular}{ccccccc}
			\\[-0.8em]
			$U_{b}$ (m s$^{-1}$) & $Re$ & $\Delta P$ (Pa) & $\tau_{w}$ (Pa) & $u_{\tau}$ (mm s$^{-1}$) & $\lambda$ (µm) & $Re_{\tau}$\\ 
			\\[-0.8em]
			\hline
			\\[-0.7em]
			0.542 & 9100 & 231 & 1.061 & 32.6 & 27.4 & 270\\
			\\[-1em]
			0.819 & 13800 & 477 & 2.188 & 46.8 & 19.1 & 390\\
			\\[-1em]
			1.094 & 18400 & 795 & 3.648 & 60.5 & 14.8 & 510\\
			\\[-1em]
			1.371 & 23000 & 1179 & 5.407 & 73.6 & 12.1 & 620\\
			\\[-1em]
			1.647 & 27700 & 1627 & 7.465 & 86.5 & 10.3 & 730\\
			\\[-1em]
			1.924 & 32300 & 2139 & 9.814 & 99.2 & 9.0 & 830\\
			\\[-1em]
			2.197 & 37000 & 2711 & 12.437 & 111.7 & 8.0 & 940\\
		\end{tabular}
	\end{ruledtabular}
\end{table*}

Measurements were conducted for seven different conditions of bulk velocity, $U_b = \dot{m}/\rho HW$, all of which are shown in table \ref{tab:1} for water. Here, $\rho$ is the fluid density. In the case of water, the Reynolds number, $Re = \rho U_{b}H/\mu_{w}$, was between the range of 4600 and 37000. The symbol, $\mu_{w}$, represents the dynamic viscosity of the fluid corresponding to the shear-rate at the wall. While this is a variable for the polymer solutions, for a Newtonian fluid such as water, the dynamic viscosity is consistently 0.89 mPa s at 25$^o$C \cite{Nagashima1977,Collings1983}. Therefore, the $Re$ of the polymer solutions are calculated later in section \ref{sec:II.D}, when the wall-shear-rates and steady-shear viscosity are obtained. The wall shear stress, $\tau_{w}$, was established using measurements of the streamwise pressure gradient, i.e. $\tau_{w} = h \Delta P/ \Delta x$, where $h = H/2$ is the half-channel height. The friction velocity, $u_{\tau} = (\tau_{w}/\rho)^{1/2}$, wall units, $\lambda = \mu_{w}/u_{\tau} \rho$, and friction Reynolds number, $Re_{\tau} = \rho u_{\tau}h/\mu_{w}$, were then subsequently determined; the results for which are listed in table \ref{tab:1} for the flow of water.

\begin{table*}[b]
	\begin{ruledtabular}
		\centering
		\caption{\label{tab:2}Flow properties for channel flow of 170 ppm XG solution.}	
		\begin{tabular}{ccccccc}
			\\[-0.8em]
			$U_{b}$ (m s$^{-1}$) & $Re$ & $\%DR$ (\%) & $\mu_{w}$ (mPa s) & $u_{\tau}$ (mm s$^{-1}$) & $\lambda$ (µm) & $Re_{\tau}$\\ 
			\\[-0.8em] 
			\hline
			\\[-0.7em]
			0.542 & 6000 & 28 & 1.348 & 29.2 & 46.2 & 160\\
			\\[-1em]
			0.819 & 10200 & 31 & 1.206 & 40.6 & 29.8 & 250\\
			\\[-1em]
			1.094 & 14200 & 32 & 1.133 & 51.4 & 22.1 & 340\\
			\\[-1em]
			1.371 & 18500 & 33 & 1.089 & 61.8 & 17.6 & 420\\
			\\[-1em]
			1.647 & 22900 & 33 & 1.059 & 72.1 & 14.7 & 510\\
			\\[-1em]
			1.924 & 27300 & 33 & 1.038 & 82.2 & 12.7 & 590\\
			\\[-1em]
			2.197 & 31700 & 33 & 1.021 & 92.3 & 11.1 & 680\\	
		\end{tabular}
	\end{ruledtabular}
\end{table*}

\subsection{Rigid polymer solution}

The rigid polymer species considered in this investigation was the polysaccharide, xanthan gum (XG) from Sigma Aldrich (CAS No. 1138-66-2). Solid XG, in powder form, was weighed using a digital scale (AB104-S, Mettler Toledo) with a 0.1 mg resolution. The powder was then gradually added to 15 l of tap water and agitated using a stand mixer (Model 1750, Arrow Engineering Mixing Products). The concentrated 15 l master solution was then left to rest overnight for approximately 12 h. The following day, the master solution was added to 100 l of moving tap water within the flow loop. This diluted the master solution to the desired concentration of 170 ppm. A 170 ppm solution of XG produced a solution of good transparancy for PIV measurements. To ensure the solution was homogeneous, the pump was operated at 1400 rpm ($U_b = 4.380$ m s$^{-1}$) for 1 h. Near the end of the 1 h duration, $\Delta P$ was marginally growing at a rate of approximately 10 Pa min$^{-1}$, about 0.1\% increase in $\Delta P$ every minute. This was considered sufficiently steady-state. After the 1 h time mark, the pump speed was reduced to 800 rpm, corresponding to $U_b$ = 2.197 m s$^{-1}$, for the first PIV measurement at the highest $Re$. The pump speed was then reduced in increments, such that PIV measurements for each flow condition listed in table \ref{tab:2} were taken. At all of the measured flow rates listed in table \ref{tab:2}, no variation in $\Delta P$ was observed during the PIV acquisition time. Therefore, any mechanical degradation or polymer de-agglomeration was likely negligible after the 1 h mixing phase. Lastly, three samples were collected for shear viscosity measurements using an access port. One sample was collected at the beginning ($U_b$ = 2.197 m s$^{-1}$), middle ($U_b$ = 1.371 m s$^{-1}$) and end ($U_b$ = 0.542 m s$^{-1}$) of data collection using PIV.

\subsection{Steady shear viscosity}\label{sec:II.B.1}

Apparent shear viscosity, $\mu$, versus shear rate, $\dot{\gamma}$, was measured for water and the 170 ppm XG solution using a torsional rheometer (HR-2, TA Instruments). Two geometries were used, a double gap concentric cylinder for low to moderate $\dot{\gamma}$ and a parallel plate for moderate to high $\dot{\gamma}$. The double gap concentric cylinder consisted of four radii: an inner cup radius (15.1 mm), an inside bob radius (16.0 mm), an outside bob radius (17.5 mm) and an outside cup radius (18.5 mm). The sample was immersed in the cup and bob at a height of 53.0 mm. The parallel plate geometry was a 60 mm diameter plate. The gap height between the plates was set to 200 µm. 

\begin{figure*}[t]
	\centering
	\includegraphics[scale=1]{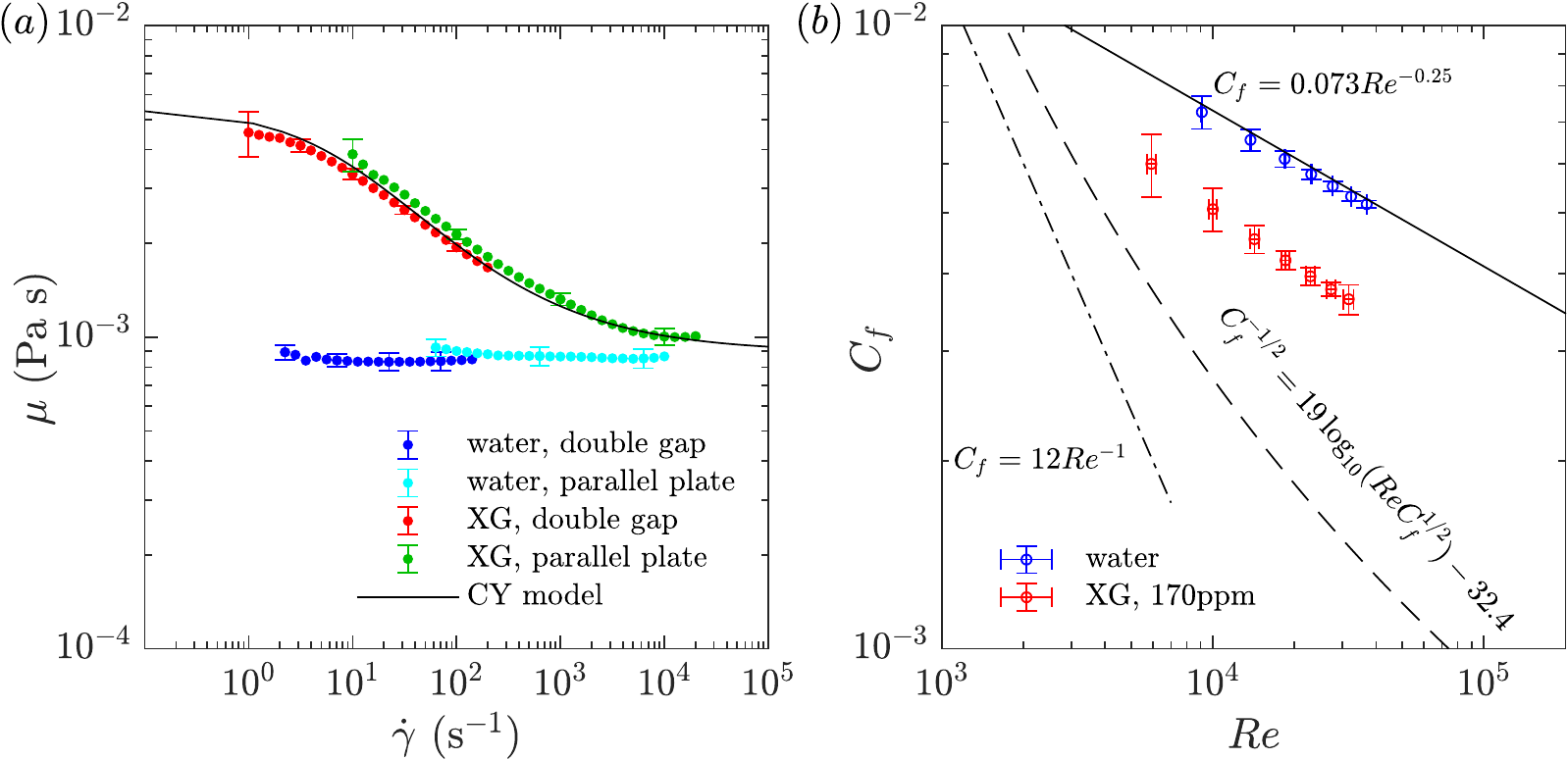}
	\caption{\label{fig:2} Steady shear viscosity measurements of 170 ppm XG solution and water. Error bars are down-sampled, showing every tenth data point and represent the range based on repeated measurements at each shear-rate.}
\end{figure*}

Figure \ref{fig:2}(\textit{a}) demonstrates measurements of $\mu$ as a function of $\dot{\gamma}$ for the 170 ppm XG solution and water at 25$^o$C. Error bars convey the range in the three repeated measurements of $\mu$ at each rotational speed and $\dot{\gamma}$. For the rigid polymer solution, measurements of $\mu$, using the double gap concentric cylinder geometry and corresponding to $\dot{\gamma} < 0.8$ s$^{-1}$, had large measurement errors due to poor sensitivity at the lower torque limit of the rheometer \cite{Ewoldt2015}. Also, measurements above 180 s$^{-1}$, using the double gap geometry had Taylor vortices that tampered the measurements. Results using the parallel plate were performed between $\dot{\gamma}$ of 10 s$^{-1}$ and 20000 s$^{-1}$ for the XG solution. For $\dot{\gamma}$ less than 10 s$^{-1}$ errors were large due to the lower torque limit of the rheometer, while for $\dot{\gamma}$ greater than 20000 s$^{-1}$ secondary flows started to obscure the results. The $\mu$ for water was measured between $\dot{\gamma}$ of 2 s$^{-1}$ and 140 s$^{-1}$ using the double gap geometry, and $\dot{\gamma}$ of 60 s$^{-1}$ and 10000 s$^{-1}$ using the parallel plate. 

The average and standard deviation in measurements of $\mu$ for water (over all values of $\dot{\gamma}$ and the thrice repeated measurements) was 0.86 mPa s $\pm$ 3.2\%. The average value of $\mu$ for water was approximately 3.5\% different than the theoretical viscosity of water at 25$^o$C, i.e. 0.89 mPa s \cite{Nagashima1977,Collings1983}. Therefore, a 3.5\% relative systematic uncertainty was approximated for all measurements of $\mu$, including measurements of the XG solution. This uncertainty propagates to other variables, including those used for inner-normalization of flow velocity.

The trend in $\mu$ as a function of $\dot{\gamma}$ for the XG solution, shown in figure \ref{fig:2}(\textit{a}), was well approximated by the Carreau-Yasuda (CY) model,

\begin{equation}\label{eqn:2}
	\frac{\mu - \mu_{\infty}}{\mu_{0}-\mu_{\infty}} = \frac{1}{(1+(\lambda_{CY}\dot{\gamma})^{a})^{n/a}},
\end{equation}

\noindent
where $\mu_{0}$ is the zero-shear-rate viscosity, $\mu_{\infty}$ is the infinite-shear-rate viscosity, $\lambda_{CY}$ is a fitting constant with dimension of time, $n$ is a dimensionless exponent and $a$ is a fitting parameter introduced by Yasuda et al. \cite{Carreau1972,Yasuda1981}. A Levenberg-Marquardt non-linear least squares method was used to fit equation \ref{eqn:2} to the measurements of $\mu$ as a function of $\dot{\gamma}$ in MATLAB. The resulting CY fit for the XG solution had a $\mu_{0}$ of 5.4 mPa s, $\mu_{\infty}$ of 0.89 mPa s, $\lambda_{CY}$ of 0.15 s, $n$ of 0.50 and $a$ of 0.81. Equation \ref{eqn:2} with these values is shown for reference in figure \ref{fig:2}(\textit{a}) by the solid black line.

\subsection{Skin-friction coefficient and drag-reduction} \label{sec:II.D}

Plots of the skin friction coefficient, $C_{f} = 2 \tau_{w}/ \rho U_b^2$, as a function of $Re$ are shown for water and XG in figure \ref{fig:2}(\textit{b}). To determine $C_f$ for the XG flows, the wall shear stress had to first be established. The $\tau_w$ of each rigid polymer flow condition was derived based on measurements of $\Delta P$. An implicit solver was used to calculate the near wall shear rate, $\dot{\gamma}_w$, corresponding to $\tau_w$ and based on the CY model. After-which, the values of $\mu_w = \tau_w / \dot{\gamma}_w$ of each XG flow was determined. Subsequently, the variables $Re$, $u_{\tau}$, $\lambda$, and $Re_{\tau}$ were obtained, all of which are listed in table \ref{tab:2} for the rigid polymer flows. The resulting values of $Re$ were then used in plots of $C_f$ shown in figure \ref{fig:2}(\textit{b}). Error bars in the data points of $C_f$ as a function of $Re$, propagate from random errors in measurements of $U_b$ and $\Delta P$, as well as the assumed uncertainty in $\mu_w$ determined in the previous section. 

Measurements of $C_f$ for water and XG show consistency with previous investigations. The equation $C_f = 0.073 Re^{-0.25}$, shown at the top of figure \ref{fig:2}(\textit{b}), is the empirical correlation relating $C_f$ and $Re$ for 2D Newtonian turbulent channel flows prescribed by Dean et al. \cite{Dean1978}. The current measurements of $C_f$ for water agree well with the equation derived by Dean et al. \cite{Dean1978} and are within 5\% of the $C_f$ power law equation. The lower equation shown in figure \ref{fig:2}(\textit{b}) is the Virk MDR asymptote, $C_f = 19 \log_{10} (Re C_f^{1/2}) - 32.4$ \cite{Virk1970}. The measurements of $C_f$ for the XG flows are between the $C_f$ correlations of Dean et al. \cite{Dean1978} and Virk et al. \cite{Virk1970}. Therefore, the XG flows do exhibit DR; however, none of the drag-reduced flows are at MDR. The $C_f$ measurements for XG also reasonably agree with the expected trend for flows of Type B drag-reducing additives with increasing \textit{Re}. Virk \& Wagger \cite{Virk1990} detailed that Type B additives exhibit a ``ladder'' effect, where the trend in $C_f$ as a function of $Re$ would be lower, but parallel to the Newtonian $C_f$ correlation equation. In figure \ref{fig:2}(\textit{b}) a trend in $C_f$ for XG that is approximately parallel to the Dean et al. \cite{Dean1978} correlation with increasing \textit{Re} can be observed.

Drag-reduction was quantified by the attenuation in $\tau_{w}$ of the polymer solution relative to a turbulent Newtonian flow of similar $Re$. The level of attenuation in $\tau_w$ was described by the percent drag-reduction,

\begin{equation}\label{eqn:1}
	\%DR = 100\Big(1- \frac{\tau_{w,R}}{\tau_{w,N}}\Big),
\end{equation}

\noindent
where $\tau_{w,R}$ is the wall shear stress of the rigid polymer solution and $\tau_{w,N}$ is the wall shear stress of a Newtonian fluid of similar $Re$. Comparing Newtonian and non-Newtonian fluids of like $Re$ accounted for changes in the viscosity due to shear-thinning \cite{Escudier2009,Japper-Jaafar2009,Xi2019}. Measurements of water and XG were performed at the same $U_b$, and not $Re$. The $Re$ are dissimilar considering XG has a larger shear viscosity than water, as demonstrated by figure \ref{fig:2}(\textit{a}). To establish a Newtonian value of $\tau_{w,N}$ that shares a common $Re$ with the XG flows, $\tau_{w,N}$ was calculated using the equation of Dean et al. \cite{Dean1978} for $C_f$.  The $\%DR$ was then subsequently determined for each flow condition of XG, the values for which are listed in table \ref{tab:2}. As previously mentioned, the experimental measurements of $C_f$ for water was within 5\% of the Dean et al. \cite{Dean1978} skin-friction correlation. Therefore, using this equation to interpolate or extrapolate the $C_f$ measurements of water was considered an adequate approximation.

\subsection{Planar particle image velocimetry}\label{sec:2.3}

\begin{figure*}[t]
	\centering
	\includegraphics{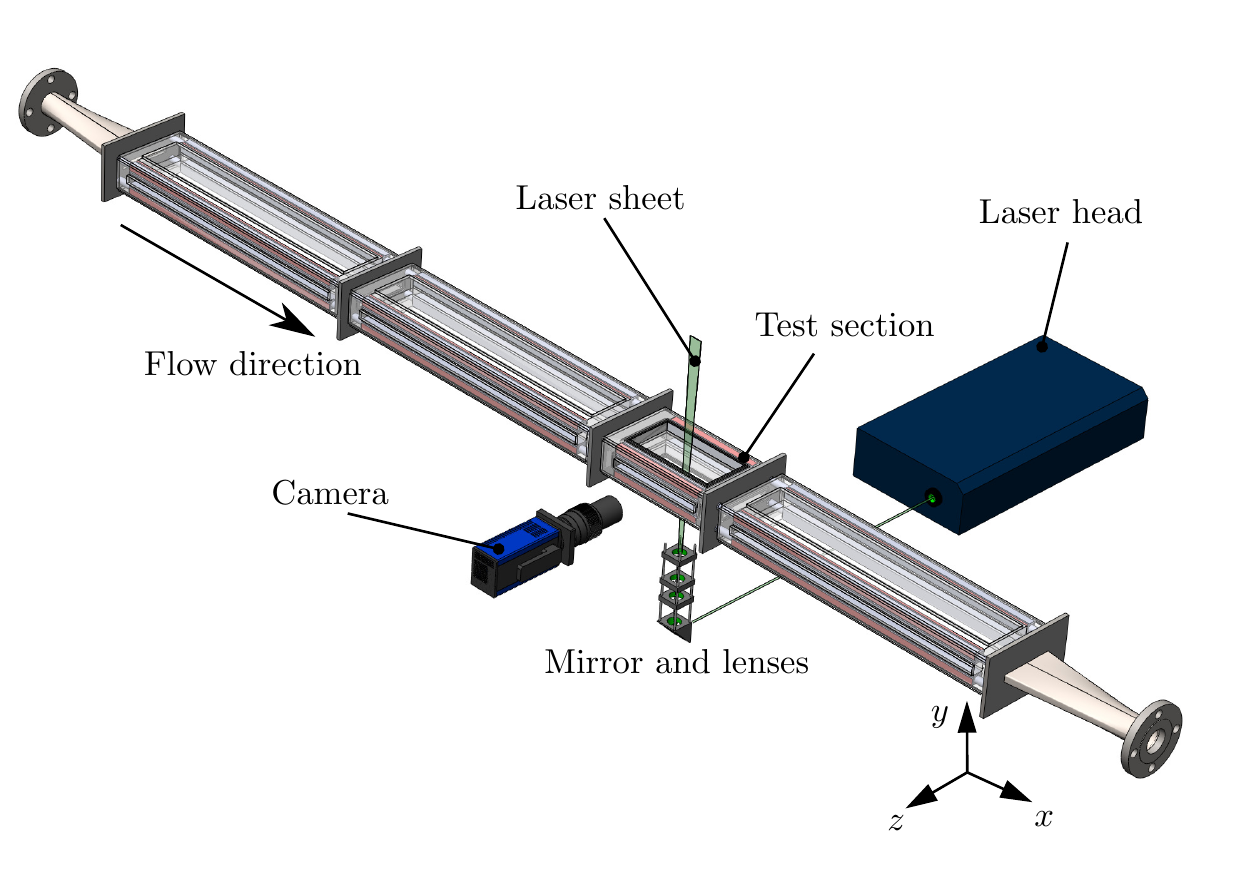}
	\caption{\label{fig:3} Isometric 3D model of the planar PIV setup relative to the glass test section and channel section.}
\end{figure*}

Planar PIV was used to characterize the velocity of the Newtonian and non-Newtonian channel flows. Images were collected using a digital camera (Imager Intense, LaVision GmbH) with a 1376 $\times$ 1040 pixels charged-coupled device (CCD) sensor. Each pixel  was 6.45 $\times$ 6.45 µm$^2$ in size with a digital resolution of 12-bit. A reduced sensor size of 1376 $\times$ 605 pixels was used to enable a higher image acquisition rate and, therefore, a faster convergence in velocity statistics. A Sigma lens with a focal length, \textit{f}, of 105 mm and an aperture size of $f/8$ was used to focus on the full height of the channel at its mid-span. The resulting magnification was 0.55, the depth-of-field was 1.30 mm, and the scaling factor was 11.81 µm pixel$^{-1}$. Figure \ref{fig:3} demonstrates the flow measurement setup relative to the test section. The camera was arranged in a portrait orientation, such that the 1376 pixel dimension of the sensor was parallel with the height of the channel. Therefore, the field of view (FOV) of the images was ($\Delta x$, $\Delta y$) =  12.28 $\times$ 16.25 mm$^{2}$. Along the \textit{x} direction, the centre of the FOV was placed at the centre of the glass test section, which is 107\textit{H} downstream of the channel inlet. 

The illumination source for the planar PIV measurements was a 90 mJ pulse$^{-1}$ Nd:YAG laser (Gemini PIV 30, New Wave Research Inc.). Two spherical lenses (one concave, the other convex) and one concave cylindrical lens expanded the 4.5 mm diameter beam output from the laser head, into a 20 mm wide (along the \textit{x}-direction) by 1 mm thick (along the \textit{z}-direction) laser sheet, at the measurement location. Silver coated hollow glass spheres, 2 µm in diameter were used to seed the flows (SG02S40 Potters Industries). Das \& Ghaemi \cite{Das2021} demonstrated that these small silver coated particles have strong side scattering and relatively consistent sizing. Synchronization between the camera and the laser was achieved using a programmable timing unit (PTU 9, LaVision GmbH) and DaVis 7.3 software (LaVision GmbH). One data set consisted of 9000 pairs of double-frame images, recorded at an acquisition rate of 7.4 Hz. The time delay, $\Delta t$, between image frames was 50 µs to 400 µs depending on the \textit{Re} of the flow. The specific value of $\Delta t$ was chosen such that the maximum particle displacement between image frames was approximately 12 pixel.

All PIV processing was performed using DaVis 8.4 software (LaVision GmbH). First, the minimum intensity of all images was subtracted from each image. Next, each data set is normalized with their respective average ensemble intensity. The instantaneous velocity along the streamwise and wall-normal directions were defined as \textit{U} and \textit{V}, respectively. Angle brackets, $\langle ... \rangle$, were used to denote the ensemble average of the variables over time and the \textit{x}-direction. The latter averaging is applied due to the homogeneity of the fully developed turbulent channel flows in the streamwise direction. Fluctuations in the streamwise and wall-normal velocity were denoted as $u$ and $v$. High spatial resolution profiles of mean streamwise velocity, $\langle U \rangle$, were established using the ensemble-of-correlation method with a final interrogation window (IW) size of 6 $\times$ 6 pixels (0.07 $\times$ 0.07 mm$^2$) and 83\% overlap between neighboring IWs \cite{Kahler2012}. The resulting profiles of $\langle U \rangle$ had a single pixel spatial resolution (0.3$\lambda$-1.5$\lambda$, depending on \textit{Re}). The lower limit of the measurements in $\langle U \rangle$ was $y = 35$ µm, which corresponds to $y^+ = 0.76$ to 3.15, depending on \textit{Re}. The instantaneous velocities, $U$ and $V$, were determined using a multi-pass cross-correlation algorithm with an initial IW size of 64 $\times$ 64 pixels and a final IW size of 32 $\times$ 32 pixels (0.38 $\times$ 0.38 mm$^2$), both with 75\% overlap between adjacent IWs. The spatial resolution of instantaneous velocity measurements was 8 pixel or 0.09 mm (2$\lambda$-12$\lambda$). Vector post-processing using the universal outlier detection algorithm developed by Westerweel \& Scarano \cite{Westerweel2005} was used to remove any spurious vectors in the measurements of \textit{U} and \textit{V}. After which, the Reynolds normal stresses, $\langle u^2 \rangle$ and $\langle v^2 \rangle$, and the Reynolds shear stress, $\langle uv \rangle$, were determined.

The wall location was determined based on the local intensity maxima, $I_{max}$, that forms due to the glare line of the wall in the average intensity distribution of the PIV images. The uncertainty in the wall location was considered to be the extent of the high intensity glare, which was assumed to be the $\Delta y$ separating $I_{max}$ and $I_{max}/e^2$ \cite{AbuRowin2017}. The corresponding uncertainty in the wall location was estimated to be approximately 3 pixel or 35.4 µm (0.8$\lambda$ - 4.4$\lambda$). Errors in the wall location were treated as an uncertainty in $y$ and are a contributing factor to the error bars in wall-normal distributions of mean velocity and Reynolds stresses.

Variables scaled using inner normalization were identified with the superscript +. Velocity statistics were normalized with the friction velocity, $u_{\tau} = (\tau_w/\rho)^{1/2}$, while positional coordinates were normalized with the wall units, $\lambda = \mu_w / u_{\tau} \rho$, as listed in tables \ref{tab:1} and \ref{tab:2}. Error propagation was used to derive the uncertainties in $u_{\tau}$ and $\lambda$ based on the assumed errors in $\mu_w$ (see section \ref{sec:II.B.1}) and random errors in $\Delta P$. A conservative 0.1 pixel uncertainty in the PIV measurements of $U$ and $V$ was also assumed \cite{Raffel2018}. Such uncertainties in the inner scaling variables and the velocity measurements were reflected by error bars in plots of $\langle U \rangle^+$, $\langle u^2 \rangle^+$, $\langle v^2 \rangle^+$, and $\langle uv \rangle^+$.

An important facet of our analysis involved a spatial gradient in the mean velocity profile along $y$, i.e. d$\langle U \rangle$/d$y$. To remove high frequency experimental noise and to differentiate the profile, a moving second-order polynomial filter was applied to the distribution of $\langle U \rangle$ with respect to $y$. The length of the filter was 24 pixel or 283 µm (6$\lambda$-35$\lambda$, depending on \textit{Re}). Coefficients of the fitted second-order polynomial were used to calculate $\text{d} \langle U \rangle / \text{d} y$, and then established the indicator function, $\zeta = y^+ \text{d}\langle U \rangle / \text{d} y^+$. We also made an attempt at determining the mean viscosity profile, $\langle \mu \rangle$, of the non-Newtonian flows by evaluating the CY model for each wall-normal gradient in mean velocity, i.e. $\text{d} \langle U \rangle / \text{d} y$. This is an assumption; one that is rather bold for a turbulent flow. As such, we denote the profiles as a ``pseudo-mean viscosity," and is indicated by $\tilde{\mu}$.

\section{Results}\label{sec:3}

\subsection{Newtonian turbulent channel flow}\label{sec:3.1}

\begin{figure}[b]
	\centering
	\includegraphics[scale=1]{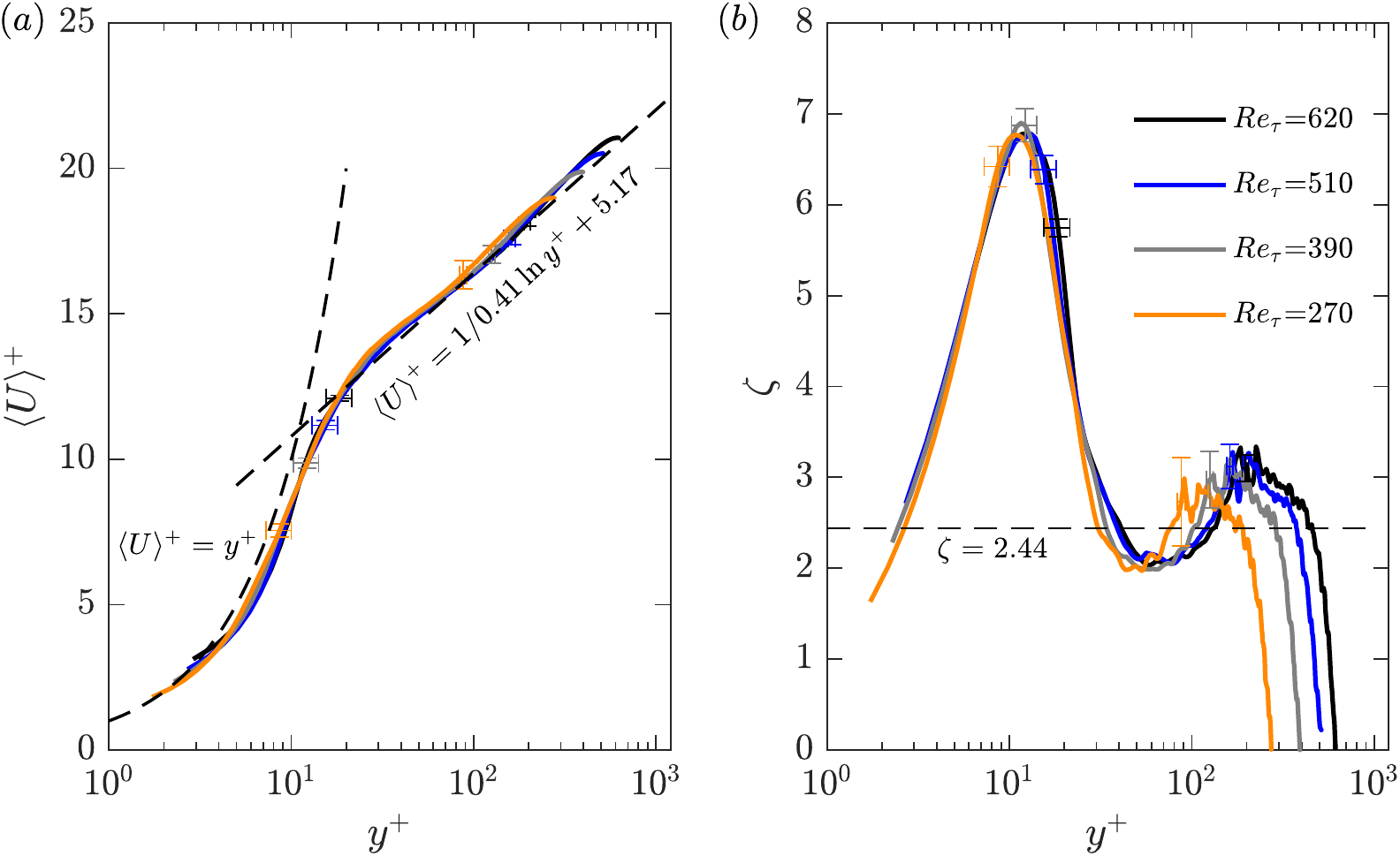}
	\caption{\label{fig:4} Inner-normalized distributions of (a) mean streamwise velocity and (b) the indicator function, for Newtonian flows.}
\end{figure}

\begin{figure}[b]
	\centering
	\includegraphics[scale=1]{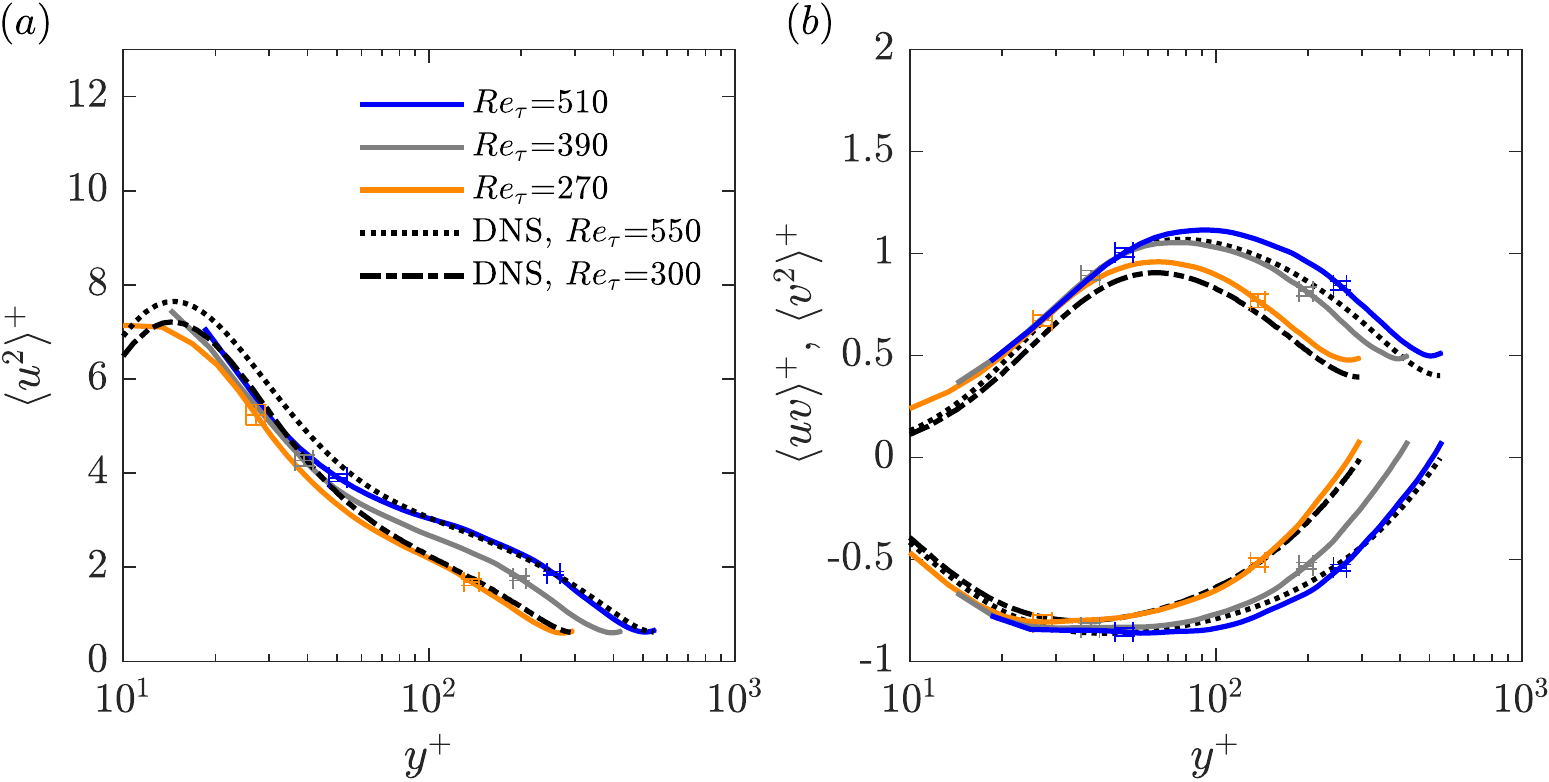}
	\caption{\label{fig:5} Inner normalized profiles of (a) streamwise Reynolds stress, (b) wall-normal and Reynolds shear stresses, for Newtonian flows.}
\end{figure}

The following section begins by comparing measurements of the mean velocity profiles for water with the Newtonian law of the wall in figure \ref{fig:4}. For brevity, only experimental data for water with a $Re_{\tau}$ less than or equal to 620 are plotted. These conditions of $Re_{\tau}$ were chosen because they are similar in magnitude to the $Re_{\tau}$ conditions of the XG flows listed in table \ref{tab:2}. Following the plots of $\langle U \rangle^+$, measurements of the Reynolds stresses for water are shown in figure \ref{fig:5}. Three experimental Reynolds stress profiles with $Re_{\tau}$ of 270, 390 and 510 are presented on the same axes as the Reynolds stresses derived from Newtonian channel flow DNS by Iwamoto et al. \cite{Iwamoto2002} at $Re_{\tau}=300$ and Lee \& Moser \cite{Lee2015} at $Re_{\tau}=550$. The error bars in figures \ref{fig:4} and \ref{fig:5} are a result of uncertainties propagating from $\mu$, $\Delta P$, $U$ and $y$. For clarity, only two error bars are shown for each profile, one approximately in the buffer layer, the other within the outer layer.

Figure \ref{fig:4}(\textit{a}) demonstrates that all experimental profiles of water show good agreement with the law of the wall. The profiles were limited to $y > 35$ µm, which corresponds to $y^+ = 1.29$ to 2.89, for $Re_{\tau}$ between 270 and 620. For $y^+ < 5$ and greater than their respective lower limit, experimental measurements show a relative overlap with the profile of the linear viscous sublayer, $\langle U \rangle^+ = y^+$. Farther from the wall, all of the experimental distributions in figure \ref{fig:4}(\textit{a}) overlap with the log law, $\langle U \rangle^+ = 1/\kappa \ln y^++B$. A Von K\'arm\'an constant, $\kappa$, of 0.41 and intercept, $B$, of 5.17, as prescribed by Dean et al. \cite{Dean1978} for 2D Newtonian channel flows, is shown for comparison. Distributions of $\zeta$ shown in figure \ref{fig:4}(\textit{b}) accentuate the logarithmic dependence of $\langle U \rangle^+$ with respect to $y^+$. The profiles of $\zeta$ imply that $\kappa$ is larger than 0.41 in the logarithmic layer for all profiles of water. Comparing the experimental profiles of $\langle U \rangle^+$ for different $Re_{\tau}$, all distributions appear to overlap with one another within the boundaries of measurement uncertainties. The DNS of a Newtonian channel flow by Lee \& Moser \cite{Lee2015} demonstrated that profiles of $\langle U \rangle^+$ over a wider $Re_{\tau}$ range of 180 to 5000 also overlapped. The current experimental results for water also reflect universality in their distributions of $\langle U \rangle^+$ and $\zeta$ among different $Re_{\tau}$.

Figure \ref{fig:5}(\textit{a}) presents experimental profiles of $\langle u^2 \rangle^+$ relative to Newtonian channel flow DNS. For water with a $Re_{\tau}=510$, instantaneous PIV measurements with IWs of 32 $\times$ 32 pixels and 75\% overlap, translates to a spatial resolution of 7.8$\lambda$. As a result, the linear viscous sublayer and a portion of the buffer layer is missed in these measurements. However, for lower $Re_{\tau}$ the spatial resolution of the measurements improve. The scenario with $Re_{\tau}=270$ has a spatial resolution of 3.5$\lambda$ and has measurements that extend to wall-normal locations as small as $y^+ = 9$. Within the logarithmic and outer layers, the experimental results overlap with their DNS counterparts at similar $Re_{\tau}$. The moderate $Re_{\tau} = 390$ case demonstrates consistency, considering it lies between the two DNS and experimental profiles at lower and higher $Re_{\tau}$. Figure \ref{fig:5}(\textit{b}) shows experimental and DNS profiles of $\langle v^2 \rangle^+$ and $\langle uv \rangle^+$. Similar to the distributions in $\langle u^2 \rangle^+$, experimental profiles in $\langle v^2 \rangle^+$ and $\langle uv \rangle^+$ agree well with the DNS results at similar $Re_{\tau}$. However, there are some small discrepancies. For example, the experimental profile of $\langle v^2 \rangle^+$ at $Re_{\tau}=510$ appears to be minutely larger than the DNS profile of $\langle v^2 \rangle^+$ at $Re_{\tau} = 550$ for $y^+>100$. Overall, the experimental mean velocity and Reynolds stress measurements show consistency and agreement with 2D Newtonian channel flow DNS. Therefore, we can proceed to the results of the non-Newtonian solution with relatively good confidence in the validity of the measurements. It should also be noted that the spatial resolution of the measurements will improve with the addition of polymers, considering $\%DR$ is coupled with a reduction in $u_{\tau}$ and an increase in $\lambda$. This can be observed by comparing the larger values of $\lambda$ for XG flows with the $\lambda$ values of water in tables \ref{tab:1} and \ref{tab:2}.

\subsection{Non-Newtonian turbulent channel flow}\label{sec:3.2}

\begin{figure}[t]
	\centering
	\includegraphics[scale=1]{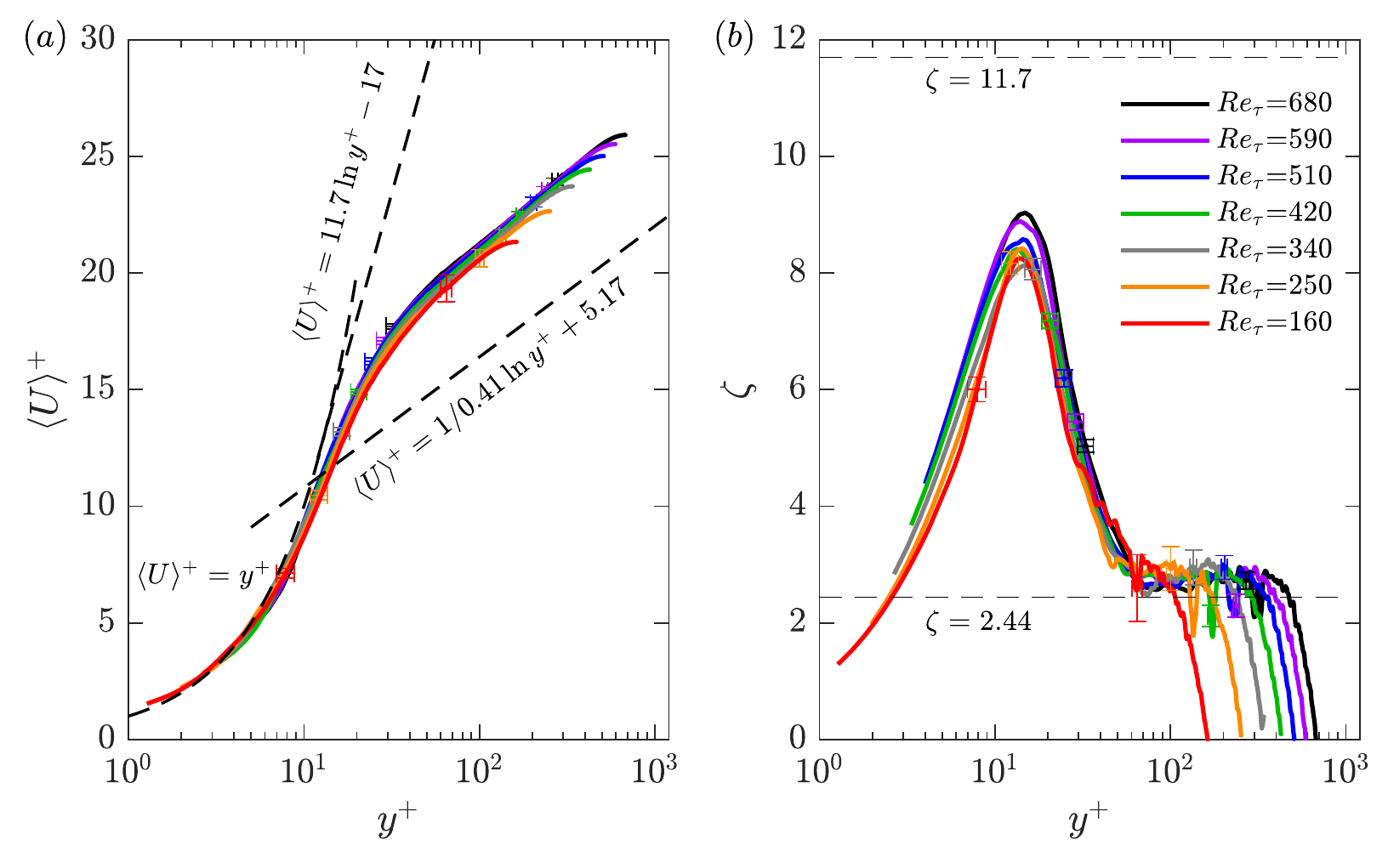}
	\caption{\label{fig:6} Inner-normalized distributions of (a) mean streamwise velocity and (b) the indicator function, for flows with 170ppm XG solution.}
\end{figure}

\begin{figure}[b]
	\centering
	\includegraphics[scale=1]{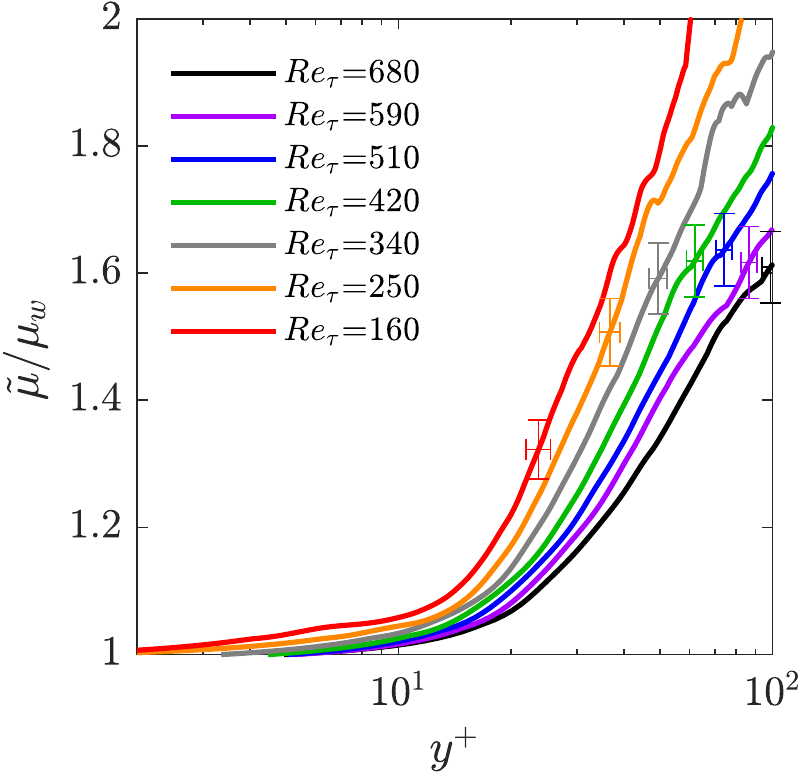}
	\caption{\label{fig:7} The pseudo-mean viscosity normalized by wall viscosity as a function of inner-normalized wall location. Vertical lines represent the lubricating layer thickness.}
\end{figure}

\begin{figure}[b]
	\centering
	\includegraphics[scale=1]{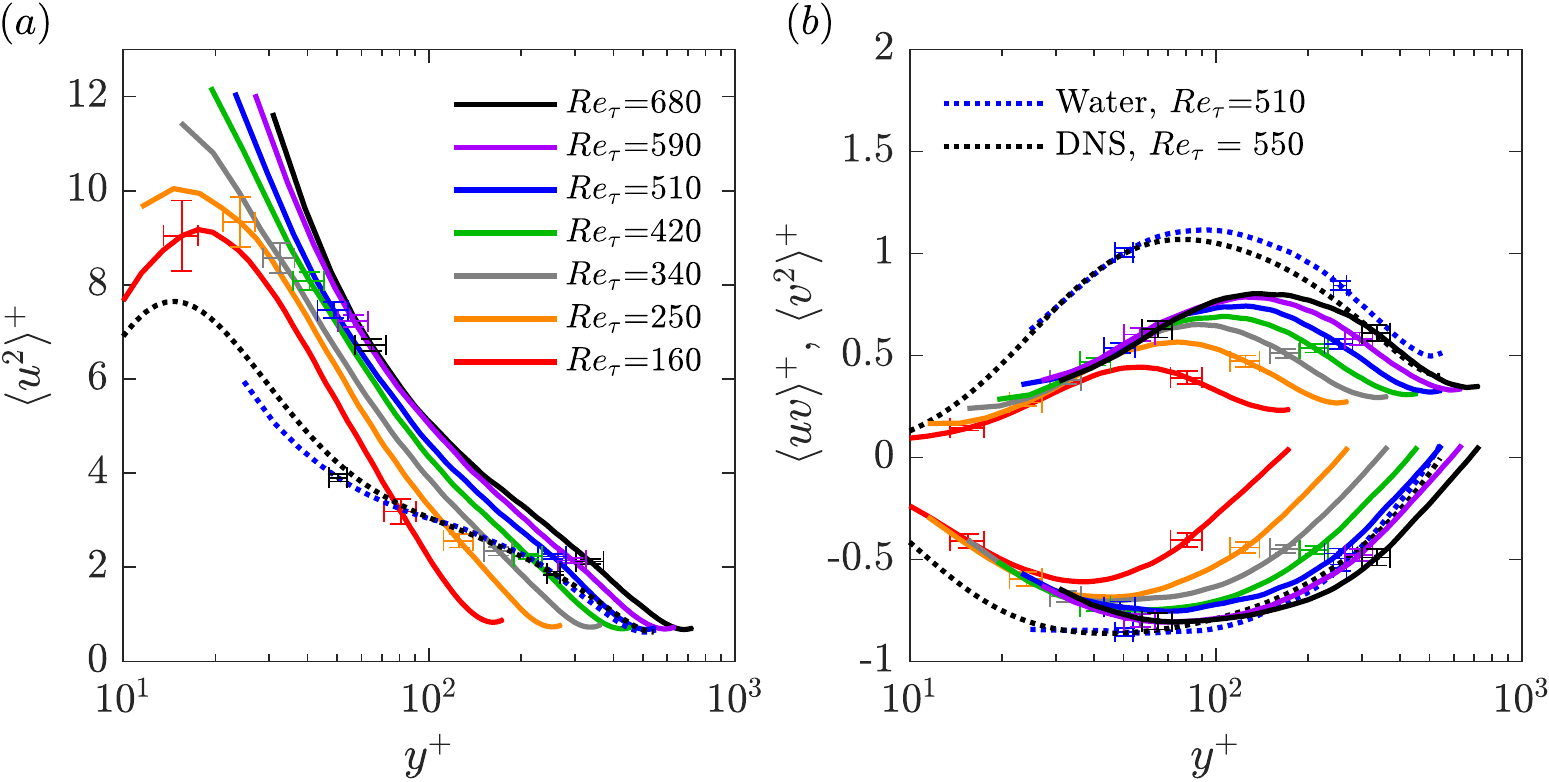}
	\caption{\label{fig:8} Inner normalized profiles of (a) streamwise Reynolds stress, (b) wall-normal and Reynolds shear stresses, for flows with 170ppm XG solution.}
\end{figure}

The current section investigates the turbulent flow of the XG solution with varying $Re_{\tau}$. Figure \ref{fig:6} presents distributions of $\langle U \rangle^+$ and $\zeta$, figure \ref{fig:7} shows profiles of $\tilde{\mu}$ with respect to $y^+$, and lastly, figure \ref{fig:8} demonstrates the Reynolds stresses. We refer the reader to table \ref{tab:2} for details pertaining to the different flow scenarios. Results of the mean velocity profile are discussed first.

Profiles of $\langle U \rangle^+$ for the XG scenarios are shown in figure \ref{fig:6}(\textit{a}). Near the wall, experimental distributions of $\langle U \rangle^+$ conform well with the linear viscous sublayer profile, $y^+ = \langle U \rangle^+$, for all $Re_{\tau}$ under consideration. The upper limit of the linear viscous sublayer appears to grow relative to Newtonian wall turbulence. For a Newtonian turbulent channel flow, the linear approximation of the viscous sublayer is valid to within 10\% at $y^+  = 5$ \cite{Pope2000}. If a 10\% confidence interval from $y^+ = \langle U \rangle^+$ is used as a threshold, we can approximate the size of the linear viscous sublayer for the non-Newtonian profiles shown in figure \ref{fig:6}(\textit{a}). The following table \ref{tab:3} lists the size of the linear viscous sublayer for the flows, both in inner- and outer-scaling. The size in inner-scaling is denoted, $y_v^+$, while the size in outer-normalization is $y_v/h$. All values of $y_v^+$ are between 8 and 12, demonstrating that the linear viscous sublayer is expanded relative to Newtonian wall turbulence, which has a $y_v^+$ between 3 and 5 \cite{Pope2000}. With increasing $Re_{\tau}$, the non-Newtonian values of $y_v^+$ increase subtly, implying that the very near wall profiles might be slightly different, and potentially depend on the small increase in $\%DR$ with increasing $Re_{\tau}$, as shown in table \ref{tab:2}. However, with error bars, these differences could be a result of uncertainty in the measurements. At the larger $Re_{\tau}$, between 510 and 680, the linear sublayer appears to saturate and nearly approach the tri-section point, $(y^+, \langle U \rangle^+) = (11.6,11.6)$, where the Virk MDR asymptote, $\langle U \rangle^+ = 11.7 \ln y^+ -17$, intersects with $y^+ = \langle U \rangle^+$ and the Newtonian log law. Values of the outer-scaled thicknesses, $y_v/h$, decrease with increasing $Re_{\tau}$, mainly due to the large shrinkage in $y_v$ caused by increasing $Re_{\tau}$.

\begin{table*}[h!]
	\begin{ruledtabular}
		\centering
		\caption{\label{tab:3} Linear viscous sublayer sizes for non-Newtonian flows in inner- and outer-scaling.}	
		\begin{tabular}{ccc}
			\\[-0.8em]
			$Re_{\tau}$ & $y_v^+$ & $y_v/h$\\ 
			\\[-0.8em] 
			\hline
			\\[-0.7em]
			160 & 8.3 & 0.051 \\
			\\[-1em]
			250 & 9.1 & 0.036\\
			\\[-1em]
			340 & 10.1 & 0.030\\
			\\[-1em]
			420 & 10.8 & 0.025\\
			\\[-1em]
			510 & 11.5 & 0.023\\
			\\[-1em]
			590 & 11.2 & 0.019\\
			\\[-1em]
			680 & 11.4 & 0.017\\	
		\end{tabular}
	\end{ruledtabular}
\end{table*}

Farther from the wall at $y^+ > 30$, figure \ref{fig:6}(\textit{a}) demonstrates a larger $\langle U \rangle^+$ relative to the logarithmic law of the wall; an observation common for drag-reduced flows. Virk \cite{Virk1971}, and later Warholic et al. \cite{Warholic1999}, demonstrated that LDR flows form a Newtonian plug profile, which is observed as an increase in the log law intercept, $B$, but a similar $\kappa$, relative to the log law distribution of a Newtonian fluid. Virk \cite{Virk1971} detailed that the growth in $B$ was proportional with $\%DR$. A larger $\%DR$ would result in an increased buffer layer thickness (deemed the elastic sublayer) and hence an enhancement in $B$. Findings from Warholic et al. \cite{Warholic1999} showed that a Newtonian plug exists only for LDR flows with $\%DR < 35\%$. Given $\%DR$ of the present XG flows are between 28-33\% (see table \ref{tab:2}), the current XG flows satisfy the criteria for LDR. Therefore, our measurements agree well with previous observations of $\langle U \rangle^+$ profiles for polymer drag-reduced LDR flows. Furthermore, figure \ref{fig:6}(\textit{a}) demonstrates that profiles of $\langle U \rangle^+$ for XG have little dependence on $Re_{\tau}$. There is perhaps a subtle increase in $B$ for $160 < Re_{\tau} < 420$; however, this could be attributed to the small growth in $\%DR$ with increasing $Re_{\tau}$. The uncertainty in the flow measurements, shown by the error bars, also captures the small variations in $B$.

White et al. \cite{White2012} re-evaluated the efficacy of the Virk \cite{Virk1971} elastic sublayer model using the indicator function, $\zeta$, which highlights regions of strong logarithmic dependence. They compared mean velocity profiles from various experimental and numerical investigations of different $\%DR$, canonical flows and $Re$. For LDR flows, White et al. \cite{White2012} observed constant $\zeta$ (generally for $y^+ > 50$), which is indicative of a Newtonian plug. Profiles of $\zeta$ shown in figure \ref{fig:6}(\textit{b}) also demonstrate regions of constant $\zeta$, providing further evidence of a Newtonian plug for rigid polymer solutions. For all $Re_{\tau}$, these regions of constant $\zeta$ are observed for $y^+ > 60$. This lower limit of $y^+ = 60$ is larger than the lower limit of $y^+ = 30$ for the Newtonian log layer \cite{Pope2000}, demonstrating an expansion of the viscous sublayer. The peak values of $\zeta$ for XG at $y^+=15$ is greater than the peak values of $\zeta$ for water as seen in figure \ref{fig:4}(\textit{b}). The implication is that the slope of $\langle U \rangle^+$ within the buffer layer is larger for flows of XG relative to water. A larger slope in $\langle U \rangle^+$ is indicative of an ``effective slip" in the buffer layer which, in turn, results in an increase in $\langle U \rangle^+$ within the logarithmic layer \cite{Lumley1969,Virk1971}. Another observation is that the constant value of $\zeta$ for the XG flows in the Newtonian plug layer, are marginally larger than the values of $\zeta$ observed for water in the logarithmic layer shown in figure \ref{fig:4}(\textit{b}). This indicates that $\kappa$ may decrease slightly with increasing $\%DR$; something also noted by White et al. \cite{White2012} for LDR flows. White et al. \cite{White2012,White2018} broadly suggested that the inner-normalized mean velocity profile of a polymer drag-reduced flow depends on the Reynolds number, polymeric properties and the canonical flow. We demonstrate that if $\%DR$ is constant, distributions of the inner-normalized mean velocity profiles of a rigid polymer solution are relatively independent of $Re_{\tau}$ within the inner layer of the flow.

Figure \ref{fig:7} demonstrates distributions of the normalized pseudo-viscosity, $\tilde{\mu}/\mu_w$, with respect to $y^+$, for the XG flows of different $Re_{\tau}$. The profiles of $\tilde{\mu}$ are an approximation of the mean viscosity in the near-wall region. Intuitively, the decreasing trend in $\tilde{\mu}$ with increasing $Re_{\tau}$ at a given $y^+$ is plausible. Flows of higher $Re_{\tau}$ have larger $\text{d} U / \text{d} y$, hence $\tilde{\mu}$ should be correspondingly lower relative to a flow of smaller $Re_{\tau}$. For $y^+<10$, all XG flows have distributions of $\tilde{\mu}/\mu_w$ that are approximately constant; only growing subtly by about 1\% with increasing $y^+$. As $y^+$ increases beyond 10 all profiles experience a dramatic increase in the magnitude of $\tilde{\mu}/\mu_w$. The precise $y^+$ location where this inflection in $\tilde{\mu}/\mu_w$ occurs depends on the $Re_{\tau}$ being considered. The thickness of the near-wall region of approximately constant $\tilde{\mu}$ appears to conform well with the peak in profiles of $\zeta$, shown in figure \ref{fig:6}(\textit{b}) and indicative of the central location of the buffer layer. The inner-normalized thickness of this region of constant $\tilde{\mu}$ grows with increasing $Re_{\tau}$. However, the value of $\tilde{\mu}/\mu_w$, appears to monotonically decrease with increasing $Re_{\tau}$ at any chosen value of $y^+$. Flows of large $Re_{\tau}$ experience a less aggressive change in $\tilde{\mu}$ with respect to $y^+$, but the size of their near-wall region of low viscosity is larger. Generally, all flows experience a large and sudden change in $\tilde{\mu}$ for $y^+$ between of 10 and 30. For example, the XG flow with $Re_{\tau} = 160$ has a $\tilde{\mu}$ that is 50\% larger than $\mu_w$ at $y^+ = 30$. A near wall region of constant mean viscosity that suddenly and dramatically increases with respect to $y^+$ has also been observed from numerical simulations using GN models \cite{Singh2017,Singh2018,Arosemena2020,Arosemena2021}. Our results appear to qualitatively agree with the results of DNS using in-elastic shear-thinning GN models \cite{Singh2017,Singh2018,Arosemena2020}. This is despite the approximation used to derive the pseudo-viscosity profile, $\tilde{\mu}$, based on 2D velocity data.

Figure \ref{fig:8}(\textit{a}) presents plots of $\langle u^2 \rangle^+$ for the XG flows alongside experimental data of water with $Re_{\tau} = 510$ and Newtonian channel flow DNS from Lee \& Moser \cite{Lee2015} with $Re_{\tau}=550$. Unlike the experimental results for water shown in figure \ref{fig:5}(\textit{a}), the peak in $\langle u^2 \rangle^+$ could be resolved for at least the two lowest $Re_{\tau}$ scenarios, i.e. $Re_{\tau} = 160$ and 250. The use of XG makes resolving the peak in $\langle u^2 \rangle^+$ easier, since drag-reducing additives have been shown to shift the peak in $\langle u^2 \rangle^+$ farther from the wall relative to Newtonian fluids \cite{Warholic1999, Escudier2009}. In general, the magnitude in $\langle u^2 \rangle^+$ for all $Re_{\tau}$ scenarios is  increased relative to the experimental profile for water shown in figure \ref{fig:8}(\textit{a}). The amount by which the XG profile of $\langle u^2 \rangle^+$ increases depends on the $Re_{\tau}$ being considered. For example, comparing XG and water at similar $Re_{\tau}$ of 510, the XG profile of $\langle u^2 \rangle^+$ is larger for nearly all $y^+$.

Profiles of $\langle v^2 \rangle^+$ are the positive distributions shown in figure \ref{fig:8}(\textit{b}). Relative to Newtonian profiles of similar $Re_{\tau}$, distributions of $\langle v^2 \rangle^+$ for the XG solutions demonstrate significant attenuation along all values of $y^+$. This can easily be seen by comparing the plots of $\langle v^2 \rangle^+$ for XG at $Re_{\tau} = 510$ with the experimental profile of water at $Re_{\tau}=510$. Distributions of $\langle uv \rangle^+$ correspond to the negative profiles shown in figure \ref{fig:8}(\textit{b}). Unlike $\langle v^2 \rangle^+$, profiles of $\langle uv \rangle^+$ are only strongly attenuated near the wall, relative to Newtonian distributions of comparable $Re_{\tau}$. The values of $\langle uv \rangle^+$ are similar for $y^+ > 150$ when comparing XG and water at a $Re_{\tau}$ of 510. While for $y^+<150$, the XG solution shows a large reduction in the magnitude of $\langle uv \rangle^+$, when contrasted with the profile of water with a similar $Re_{\tau}$ of 510. Therefore, relative to Newtonian profiles of similar $Re_{\tau}$, solutions of XG at LDR exhibit strong attenuation in $\langle v^2 \rangle^+$ throughout the complete half-channel; however attenuation in $\langle uv \rangle^+$ is confined to a portion of the channel near the wall. Comparing the Reynolds stress profiles of XG with one another, all distributions for XG shown in figure \ref{fig:8} increase in magnitude monotonically with increasing $Re_{\tau}$ at a given $y^+$, similar to the trend in the Reynolds stresses for Newtonian fluids of increasing $Re_{\tau}$. 

In summary, the rigid polymer solution demonstrates larger profiles in $\langle U \rangle^+$ within the logarithmic layer relative to water, conducive of a Newtonian plug. Non-Newtonian flows of different $Re_{\tau}$ and similar $\%DR$ had overlapping profiles in $\langle U \rangle^+$, within the margin of measurement uncertainty. When compared to experiments of Newtonian turbulence at a similar $Re_{\tau}$, XG exhibits larger profiles in $\langle u^2 \rangle^+$, and smaller profiles in $\langle v^2 \rangle^+$, for all $y^+$. Attenuation in $\langle uv \rangle^+$ is observable, but only near the wall. These findings share similarities with numerical investigations using in-elastic models, such as the GN power-law or Carreau constitutive equations. Singh et al. \cite{Singh2018} used a power-law model to simulate an in-elastic non-Newtonian turbulent pipe flow of $Re_{\tau}$ between 323 and 750. Constant material properties were maintained across their different cases of $Re_{\tau}$ to evaluate the effect of Re on the flow statistics, much like what is demonstrated in the present experimental investigation. Singh et al. \cite{Singh2018} observed a Newtonian plug for all flow conditions, profiles of $\langle U \rangle^+$ that overlapped across different $Re_{\tau}$, an enhancement in $\langle u^2 \rangle^+$, attenuation in the radial and azimuthal Reynolds stresses, and a confined near wall attenuation in $\langle uv \rangle^+$, relative to a Newtonian flow of similar $Re_{\tau}$. Contrasting this with experiments using flexible polymers or DNS using elastic models, such as FENE-P, the same observations can be made for mean velocity statistics of generally any LDR flow, including the current findings. Consistency in the mean velocity statistics of elastic and in-elastic DR suggests that the net effect of DR using elastic or in-elastic additives is the same, at least for flows at LDR. This is despite their dramatically different rheology and potentially unique mechanisms for mitigating drag.

\section{Discussion - lubricating layer}

The classical theories of polymer DR have insinuated that polymers interact with turbulence in a manner that quells regions of high strain and vorticity through either an enhanced extensional viscosity or elasticity \cite{Lumley1973,DeGennes1990}. Indeed, experiments with flexible polymers in isotropic homogeneous grid turbulence demonstrate suppression of the small scale turbulent eddies that correspond to regions of the flow with high extensional strain, and thus large extensional viscosities \cite{VanDoorn1999}. However, shear-thinning properties of rigid polymers work against these postulates, in that regions with large shear rates have lower viscosities, not enhanced. A comparison of isotropic turbulence using FENE-P versus in-elastic shear-thinning constitutive models could directly contrast the local instantaneous effect of flexible and rigid polymers on turbulence. We argue that the phenomenon of DR for in-elastic shear-thinning fluids is primarily attributed to a wall-normal gradient in shear viscosity induced from the wall and not local damping of eddies. This is a hypothesis that is also supported by the recent numerical investigation by Arosemena et al. \cite{Arosemena2021}, who performed a channel flow DNS using an in-elastic Carreau constitutive model and commented on the near wall turbulent structures within the flow. They surmised that the local increase in viscosity with increasing distance from the wall produces less energetic vortices and DR. Instantaneous suppression of small-scale turbulence, as theorized by Lumley \cite{Lumley1973} or de Gennes \cite{DeGennes1990}, is perceived to be more relevant to DR using flexible polymers that form viscoelastic solutions with a large extensional viscosity.

In the present investigation, considerable wall-normal gradients in the viscosity of the XG solution are inferred based on distributions of $\tilde{\mu}$ shown in figure \ref{fig:7}. A thin layer of nearly constant viscosity is observed close to the wall, corresponding to regions occupying the linear viscous sublayer and portions of the buffer layer. At larger $y^+$, figure \ref{fig:7} demonstrates that all flows experience substantial growth in the mean viscosity. This thin near-wall layer is perhaps analogous to the low viscosity lubricating layer in the DNS of Roccon et al. \cite{Roccon2019}. In this numerical investigation, a thin layer of immiscible fluid with a different viscosity was introduced in the near wall region. When the near wall region had a viscosity comparable with that of the bulk fluid, Roccon et al. \cite{Roccon2019} observed that the surface tension between the two fluids produced DR. However, for the cases where the near wall fluid had a lower viscosity, they commented that the near wall fluid acts as a lubricating layer that results in a smaller wall friction and consequently DR. In addition to this observation, there are some notable similarities with respect to the current investigation. In their DNS, Roccon et al. \cite{Roccon2019} demonstrated that the average thickness of the lubricating layer was similar to the thickness of the expanded linear viscous sublayer, $y_v/h$, in the present experimental findings for XG. The DNS by Roccon et al. \cite{Roccon2019} attained $\%DR$ of 24\% with a lubricating layer that was 0.038$h$ in thickness; a value comparable to those of $y_v/h$ for XG, which are between 0.017\textit{h} and 0.051\textit{h}, as listed in table \ref{tab:3}.

Turbulent DR using shear-thinning liquids may also share commonalities with DR using superhydrophobic surfaces. Adding micro-scale roughness to a hydrophobic material produces a thin layer of air between the liquid and the solid boundary \cite{Rothstein2010}. The air layer causes the moving liquid to ``slip", generally resulting in large quantities of DR \cite{Ling2016,AbuRowin2019}. This apparent slip of the liquid phase produces a mean velocity profile where values of $\langle U \rangle^+$ are larger for all $y^+$, but parallel to the Newtonian law of the wall; seemingly reminiscent of the Newtonian plug in polymer DR. Indeed, Lumley \cite{Lumley1969} and Virk \cite{Virk1971} have regarded the Newtonian plug for polymer DR as being an ``effective slip". The Newtonian plug is realized in a polymer drag-reduced flow when the log layer is displaced upwards to larger $\langle U \rangle^+$ \cite{Virk1971}. The Newtonian plug and the ``effective slip" were alluded to in the results pertaining to profiles of $\langle U \rangle^+$, and was realized by the large peak in $\zeta$. For rigid polymer solutions, slippage and the Newtonian plug is perhaps a manifestation of the fluids shear-thinning rheology and the near-wall lubricating layer.

\section{Conclusions}
Solutions of xanthan gum (XG) polymer have historically demonstrated little viscoelastic and extensional properties; two rheological features often attributed to polymer drag-reduction (DR). Few existing experimental investigations have demonstrated the turbulence statistics of rigid polymers in a turbulent channel flow. The primary objective of our investigation was to scrutinize the effect of varying Reynolds number (Re) on the mean velocity and Reynolds stress profiles, independent of change in DR. Our second objective was to evaluate the wall-normal gradient in the shear viscosity for drag-reduced flows of rigid polymers. 

Measurements of the mean velocity profile and Reynolds stresses for an aqueous XG solution at friction Reynolds numbers, $Re_{\tau}$, between 160 and 680 were provided. Compared to flows of similar Re, the XG solution exhibited drag-reduction percentages, $\%DR$, between 28\% and 33\% with varying $Re_{\tau}$. The XG solution reflected $\%DR$ with little dependence on the Re and skin friction coefficient values, consistent with previous observations of DR using Type B additives. A torsional rheometer equipped with a double gap concentric cylinder and a parallel plate was used to measure the apparent shear viscosity for shear rates between 1 s$^{-1}$ to 20000 s$^{-1}$, capturing both the zero- and infinite-shear-rate viscosities of the XG solution. A Carreau-Yasuda (CY) fit was used to model the apparent shear viscosity curve. 

Inner-normalized mean velocity profiles for the XG flows of different $Re_{\tau}$ approximately overlapped. This observation demonstrated that the inner-normalized mean velocity profiles are independent of $Re_{\tau}$ for a constant DR. Relative to the Newtonian law of the wall, the intercept of the log layer was considerably larger, and the slope demonstrated marginal growth (i.e. a Newtonian plug flow). Compared to Newtonian Reynolds stress profiles of similar $Re_{\tau}$, distributions for XG exhibited enhancement in streamwise Reynolds stresses and attenuation in wall-normal Reynolds stresses for all inner normalized wall-normal coordinates. Attenuation in the Reynolds shear stress was only observed near the wall. The effect of increasing $Re_{\tau}$ in the non-Newtonian flows was the same as Newtonian, i.e. the Reynolds stresses increased in the logarithmic layer monotonically with increasing $Re_{\tau}$. The modification to the first- and second-order velocity statistics reflected consistency with results obtained from DNS using elastic and in-elastic constitutive models and previous experiments with flexible polymers.

The CY model and the spatial gradient in the mean velocity profile were used to approximate the mean viscosity profiles of each drag-reduced flow. All XG flows possessed a near wall region that was thin and had a low viscosity. Fluid at wall-normal locations immediately above this region demonstrated dramatic growth in the mean viscosity. As such, we denoted this thin low viscosity layer as a ``lubricating layer," analogous to the wall-normal viscosity stratification observed in lubricated wall-bounded flows of immiscible fluids. This lubricating layer encapsulated the expanded linear viscous sublayer and portions of the buffer layer for flows of the XG solution. Its extent corresponded roughly to the peak in the indicator function, $\zeta$. Unlike the classical theories of polymer DR, we hypothesize that rigid polymer DR is largely attributed to gradients in the mean velocity coupled with the solutions shear-thinning rheology. The lubricating layer is a by-product of this interaction and a mechanism for generating an effective slip within the buffer layer.

\begin{acknowledgments}
	The authors acknowledge the support of the Natural Sciences and Engineering Research Council of Canada (grant number: RGPIN-2020-07231)
\end{acknowledgments}

\bibliography{document}

\end{document}